\documentclass{article}

\usepackage[dblblindworkshop, final]{neurips_2026}
\workshoptitle{Sim2Science: ML with Imperfect Scientific Models}

\usepackage[utf8]{inputenc} 
\usepackage[T1]{fontenc}    
\usepackage{hyperref}       
\usepackage{url}            
\usepackage{booktabs}       
\usepackage{amsfonts}       
\usepackage{nicefrac}       
\usepackage{microtype}      
\usepackage{xcolor}         
\usepackage{graphicx}
\usepackage{graphicx}
\usepackage{subcaption}
\usepackage{enumitem}
\title{Domain Adaptive Graph Neural Networks for Neutrino Experiments}

\author{%
  Hemalata Nayak\\ 
  School of Computing, Data Sciences \& Physics\\
  The College of William \& Mary\\ 
  Williamsburg, VA 23185;\\ 
  Scientific Computing Division\\ 
  Fermi National Accelerator Laboratory\\ 
  Batavia, IL 60510 \\
  \texttt{hnayak@wm.edu}
  \And
  Giuseppe Cerati\\
  Scientific Computing Division\\ 
  Fermi National Accelerator Laboratory\\ 
  Batavia, IL 60510;\\
  Consortium for Advanced Science\\
  and Engineering (CASE)\\
  University of Chicago, Chicago, IL 60637\\
  \texttt{cerati@fnal.gov}
  \And
  Aleksandra \'Ciprijanovi\'c \\
  Scientific Computing Division\\ 
  Fermi National Accelerator Laboratory\\
  Batavia, IL 60510; \\
  Department of Astronomy \& Astrophysics\\
  University of Chicago, Chicago, IL 60637;\\
  NSF-Simons AI Institute for the Sky (SkAI)\\
 Chicago, IL 60611\\
  \texttt{aleksand@fnal.gov}
  \And
  Tammy Walton \\
  Scientific Computing Division\\ 
  Fermi National Accelerator Laboratory\\ 
  Batavia, IL 60510 \\
  \texttt{twalton@fnal.gov}
}

\begin{document}
\maketitle

\begin{abstract}
Weakly interacting particles such as neutrinos require highly efficient and precise detection methods to enable searches for new physics. Deep learning, and particularly Graph Neural Networks (GNNs), has shown promise for reconstruction tasks in liquid argon time projection chamber (LArTPC) neutrino experiments. However, differences between simulations and experimental data introduce domain shifts that can degrade model performance and potentially mimic or obscure new-physics signals. Here, we demonstrate that robust learning across datasets can still be achieved by using domain adaptation during training. We introduce, to our knowledge for the first time, domain-adaptation-enabled GNNs for robust LArTPC reconstruction tasks. We use MicroBooNE open dataset as our labeled source domain, and introduce four perturbed datasets as our unlabeled target domains-- including a fixed scale variation, moderate and severe gaussian noise, and a signal variation as a function of the position in the detector. The inclusion of adversarial-based domain adaptation during training improves performance for all four target domains, without a substantial drop in performance in the source domain. Precision in the target domain was improved up to 0.19 and recall up to 0.37, depending on the variation and the network task. These results demonstrate a potential of domain adaptation to bridge the gap between simulations and experiments, and build more trustworthy models for current and future LArTPC neutrino experiments.
\end{abstract}

\section{Introduction}

Neutrinos only interact through the weak force, so physicists build massive detectors and expose them to intense beams in order to precisely measure their properties and search for new phenomena. Maximizing the efficiency is therefore a vital requirement for reconstruction methods used in these experiments. However, even more important is to reliably know these efficiencies, as biases in simulation predictions compared to the actual data observation can lead to measurement distortions that can either hide or mimic a fake new physics signal.

Machine learning (ML) has become increasingly important in neutrino physics, with convolutional neural networks (CNNs) successfully applied to neutrino interaction classification, particle identification, and event reconstruction in experiments including NOvA, DUNE, and MicroBooNE~\cite{aurisano1604convolutional, psihas2019context, abi2020neutrino, abed2022separation, adams2019deep}. However, representing measurements from liquid argon time projection chambers (LArTPCs), the primary detector technology used by most current and future neutrino experiments, as images imposes a regular grid structure on inherently sparse and irregular detector data. Graph neural networks (GNNs) provide a natural alternative by operating directly on irregular data structures and modeling relationships between detector measurements~\cite{zhou2020graph} or describing the interactions between higher-level objects~\cite{Drielsma:2021jdv}. NuGraph2 is a heterogeneous GNN-based message-passing attention architecture developed for LArTPC reconstruction, using 2D detector-plane subgraphs connected through 3D information to provide a consistent representation of neutrino interactions~\cite{aurisano2024graph}. Building on NuGraph2, NuGraph3 is under development: a hierarchical message-passing attention network that dynamically constructs higher-level graph representations from low-level detector hits using a learned metric-space embedding. 

However, all deep learning methods, including GNNs, can suffer from a drop in performance due to \textit{domain shift} i.e., a model trained on one dataset, will perform sub-optimally (or even completely fail) when applied to a different dataset, even when the differences between them are relatively small. These differences can come from imperfect simulations, mismatch between simulated and experimental data, or differences between two different experiments. Domain adaptation (DA)~\cite{Wang_2018,wilsongarrett} includes a broad group of methods aimed at solving the domain shift problem by aligning internal latent representation of the two datasets and allowing the model to perform well on both. DA has already been demonstrated to enable more robust learning in collider physics~\cite{baalouch2019sim}, neutrino experiments~\cite{PG2018}, and astrophysics~\cite{CL2023,Agarwal2024,Parul2024,sidda}. In case of GNN applications, DA has been demonstrated to successfully improve robustness of GNNs trained to infer cosmological parameters across different cosmological simulation suites~\cite{RC2023}. 
In this work, we investigate adversarial DA using a Domain-Adversarial Neural Network (DANN)~\cite{ganin2016domain} framework to improve the robustness of the NuGraph3 GNN to simulation-to-data differences. We evaluate the impact of domain shift on model performance and assess the effectiveness of adversarial DA in mitigating these effects, with the goal of developing robust ML-based reconstruction methods for future neutrino experiments such as DUNE.

\section{Methodology}

Our graph construction follows NuGraph2~\cite{aurisano2024graph}, which connects 2D detector-plane subgraphs using 3D spatial information. The broader NuGraph3 architecture (still under active development) expands upon this framework by utilizing five features per graph node: the detector plane number, the hit's wire index, the time coordinate, and the properties of its reconstructed Gaussian pulse -- the integral (representing the total charge collected on the wire), and the RMS (representing the spatial spread or diffusion of the charge). Additionally, NuGraph3 operates as a hierarchical message-passing attention network that dynamically constructs higher-level graph representations from low-level hits via a learned metric-space embedding.\footnote{The version of NuGraph3 used in this study is available at \url{https://github.com/Hemalatanayak/nugraph/releases/tag/v0.0.1}.} Contrary to NuGraph2, where all downstream decoders are applied to a single latent space, NuGraph3 allows different decoders to operate at varying representation levels. The network consists of an encoder that projects the input hit information into a latent dimension (977 parameters), a core message-passing block (141K parameters), and modular task-specific decoders.

We train NuGraph3 with the filter and semantic decoders as described in~\cite{aurisano2024graph}. While the architecture supports distributing decoders across the hierarchy, for this study, both the filter decoder (8.5K parameters) and the semantic decoder (9.0K parameters) are applied to the same latent space output from the core message-passing block. In total, the model comprises approximately 159K trainable parameters. The filter decoder performs a binary classification task to reject hits that are not part of the primary neutrino interaction using a simple binary cross-entropy loss. The semantic decoder assigns each hit to one of five particle classes, optimized using the Recall Loss~\cite{RecallLoss} to maintain a balance between precision and recall. Each class represents the particle type that originated the hit: minimum ionizing particle (\emph{MIP} i.e., muons and charged pions), highly ionizing particle (\emph{HIP} i.e., protons), particle producing an electromagnetic cascade (\emph{shower} i.e., primary electrons or photons), electron from the decay at rest of a muon (\emph{Michel}), sparse deposits from a low-energy particle (\emph{diffuse}).

To make this architecture robust against domain shifts, we integrate it into a Domain-Adversarial Neural Network (DANN) framework~\cite{ganin2016domain}. This is achieved by introducing an adversarial branch one dense layer of size 64 followed by ReLU activation function, and an output layer of size 2 trained to distinguish between source and target data using a cross-entropy loss. Crucially, this domain classifier is applied to the same latent embeddings as the filter and semantic decoders. During training, the feature extractor learns to minimize the primary task losses on the source domain while simultaneously maximizing the error of the domain classifier. This adversarial objective is enforced using a Gradient Reversal Layer (GRL) placed before the domain classifier, which multiplies gradients by a negative constant during backpropagation. We use GRL scaling coefficient $\alpha=1$. Consequently, the network is forced to learn representations that are highly discriminative for the primary physics tasks but invariant to the underlying domain, thereby mitigating performance degradation caused by simulation-to-data discrepancies.

In all of our experiments, we use labels from the source domain $\mathcal{S}$, enabling us to calculate semantic $\mathcal{L}_\mathrm{sem}^\mathcal{S}$ and filter $\mathcal{L}_\mathrm{filt}^\mathcal{S}$ losses for this dataset. The target datasets $\mathcal{T}$ mimic real detector data with altered properties compared to the source dataset. Labels in the target domain are not used, so that the target data is not directly used to train the decoders. Latent embeddings of both source and target data are forwarded to the adversarial branch, and are used to calculate DA loss $\mathcal{L}_\mathrm{DA}^\mathcal{ST}$. Hence, our model is trained using $\mathcal{L}= \mathcal{L}_\mathrm{filt}^\mathcal{S} + \mathcal{L}_\mathrm{sem}^\mathcal{S} + \mathcal{L}_\mathrm{DA}^\mathcal{ST}$. Loss functions for this multi-task training are balanced by utilising a temperature mechanism to automatically find an
optimal performance balance between all three tasks~\cite{tempscale}.
We note that because both semantic and filter decoders operate using the same latent embeddings from NuGraph3 hierarchy, we only include $\mathcal{L}_\mathrm{DA}^\mathcal{ST}$ in the semantic decoder.

\section{Data}\label{sec:data}

To train and evaluate our DA framework, we require distinct source and target domains. Both are identically partitioned into training, validation, and test sets using a 90\% : 5\% : 5\% ratio, corresponding to 19,868, 1,103, and 1,103 events, respectively. The source domain utilizes a subset of events from the MicroBooNE open datasets~\cite{abratenko_2022_7261921, abratenko_2023_8370883}, consisting of simulated inclusive ($\nu_{\mu}$) and electron neutrino ($\nu_{e}$) interactions overlaid on real data taken with the MicroBooNE detector~\cite{MicroBooNE:2016pwy} at times when no neutrino beam was present. To mimic realistic discrepancies between simulation and real data, such as scale variation effects, electronic noise, and electron diffusion miscalibration, we generated four synthetic target datasets by systematically perturbing the hit features related to the pulse shape (integral and RMS) in the source dataset\footnote{The source dataset and scripts to generate the target datasets are available at \url{https://doi.org/10.5281/zenodo.23520376}.}

\begin{itemize}[leftmargin=*]
    \item \textbf{Target 1 (Fixed Scaling):} Fixed scaling factors were applied to the integral and RMS features independently for the U, V, and Y wire planes (e.g., scaling the U-plane integral by 0.8 and RMS by 1.5). This dataset does not map to a specific detector effect but provides a robustness test against significant scale variations.
    \item \textbf{Target 2 (Moderate Gaussian Noise):} Simulates moderate electronic noise. Gaussian perturbations were added to the integral and RMS features. The standard deviation of the noise was set to 20\% of the global robust spread of each feature (calculated as half the distance between the 16th and 84th percentiles across the entire dataset). This noise level is not necessarily matching any actual variation in the MicroBooNE detector, but was chosen to provide a visible effect in the network performance.
    \item \textbf{Target 3 (Severe Gaussian Noise):} Similar to Target 2, but applies a more severe noise level equal to 40\% of the global robust feature spread.
    \item \textbf{Target 4 (Time-Dependent RMS Shift):} Simulates variations in longitudinal electron diffusion. The RMS values in the target dataset were then shifted to reproduce the behavior of a different diffusion constant. The source (target) dataset corresponds to a diffusion constant $D_L$=6.40 (3.74) cm$^2$/s. These values were used by MicroBooNE for different simulation campaigns, so this test corresponds to a realistic variation~\cite{MicroBooNE:2021icu}.
\end{itemize}

\vspace{-0.2 cm}
\section{Results and Discussion}

For all experiments, we trained two models (without and with DA), and we report precision (TP/(TP + FP)) and recall (TP/(TP + FN)), where TP, FP, and FN represent true positives, false positives, and false negatives, respectively. Prediction labels are assigned based on the category with the maximum score. We trained for 50 epochs on A100 GPUs with a batch size of 64. The optimization process utilized AdamW~\cite{loshchilov2019decoupledweightdecayregularization} with a 0.001 learning rate and a OneCycleLR scheduler~\cite{smith2018superconvergencefasttrainingneural}. Total model training time was $\sim34.6$ minutes per epoch, with an average of $\sim$7.35 seconds per batch. Notably, models trained with DA are approximately 5\% slower per epoch than the No-DA configuration.

\vspace{-0.3cm}
\begin{table}[ht!]
\caption{Precision (P) and recall (R) of the semantic and filter decoders evaluated on the source and target test sets. Due to training with the recall loss~\cite{RecallLoss}, these are by construction the same value for the semantic decoder, while for the filter decoder, we report them as (P/R). For each experiment ($\mathrm{DA}_1$–$\mathrm{DA_4}$), decoder (semantic or filter), and domain (source or target), the better-performing result between the No-DA and DA is shown in bold. Results have a statistical precision of O(1\%) or less, estimated as the standard deviation of five repetitions of the training with different random seeds.}
\label{tab:precision-recall}
\centering
\resizebox{\linewidth}{!}{%
\begin{tabular}{lcccccccc}
\hline
 & \multicolumn{2}{c}{Semantic--Source} &
   \multicolumn{2}{c}{Semantic--Target} &
   \multicolumn{2}{c}{Filter--Source} &
   \multicolumn{2}{c}{Filter--Target} \\
\cline{2-9}
Experiment &
No-DA & DA &
No-DA & DA &
No-DA(P/R) & DA(P/R) &
No-DA(P/R) & DA(P/R) \\
\hline
DA$_1$ & \textbf{0.89} & 0.87 & 0.68 & \textbf{0.83} & \textbf{0.96} / 0.97 & 0.95 / 0.97 & 0.93 / 0.56 & 0.93 / \textbf{0.93} \\

DA$_2$ & 0.90 & 0.90 & 0.86 & \textbf{0.89} & 0.96 / 0.97 & 0.96 / 0.97 & 0.96 / 0.94 & 0.96 / \textbf{0.96} \\

DA$_3$ & \textbf{0.91} & 0.88 & 0.67 & \textbf{0.86} & \textbf{0.96} / 0.97 & 0.95 / 0.97 & \textbf{0.96} / 0.80 & 0.95 / \textbf{0.95} \\

DA$_4$ & 0.89 & \textbf{0.90} & 0.86 & \textbf{0.89} & 0.96 / 0.97 & 0.96 / 0.97 & 0.97 / 0.87 & 0.97 / \textbf{0.94} \\
\hline
\end{tabular}%
}
\end{table}




\begin{figure}[htbp!]
    \centering
\includegraphics[width=0.92\linewidth,trim=20 0 10 10,clip]{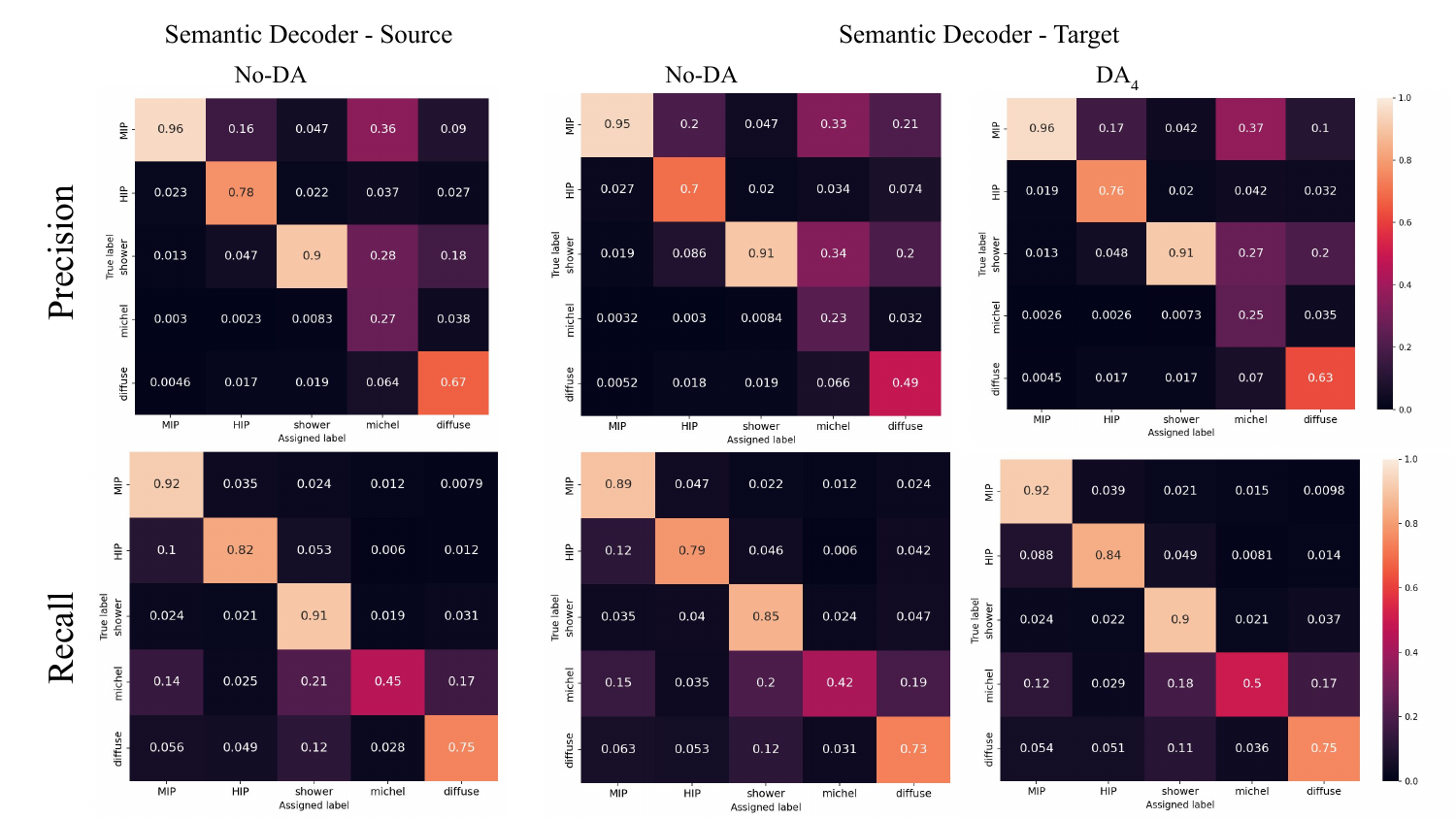}
    \caption{Confusion matrices showing precision (top) and recall (bottom) of the semantic decoder, when models are trained without (left - source domain; middle - target domain), and with DA (right - target domain). Both matrices are obtained from the same inputs: for precision the matrix is normalized by column (assigned labels), while for recall by row (true label). Results are shown for test sets of time-dependent RMS shift ($\mathrm{DA}_4$). Inclusion of DA improves both precision and recall.}\label{fig:CMs_semantic}
\end{figure}

The numbers are reported for both semantic and filter decoders when applied to source and target test datasets (Table~\ref{tab:precision-recall}). Given the small size of the datasets used in these tests, the absolute performance values are not meant to provide an ultimate measure of the network capabilities. In contrast, what is relevant is the relative comparison of such metrics across the different tests performed. The baseline (No-DA) results clearly illustrate the vulnerability of the standard architecture to domain shifts. While the baseline model achieves high precision and recall on the source test set (e.g., 0.89 for semantic and 0.96 for filter), its performance degrades significantly when evaluated on the perturbed target datasets. The degradation is most severe under fixed scaling (DA$_1$) and severe Gaussian noise (DA$_3$), where the semantic scores drop to 0.68 and 0.67, respectively. Furthermore, the filter decoder exhibits a sharp decline in recall (dropping to 0.56 in DA$_1$), indicating that the model aggressively filters out valid hits when exposed to out-of-distribution feature scales. Overall, filter precision is the least affected metric because variations typically lead to labeling more signal hits as noise, thus preserving the precision of the most represented category in the sample (signal).

Integrating the DANN framework successfully mitigates these effects across all tested perturbations. As shown in the DA columns for the target sets, the adversarial training recovers the lost performance substantially. For instance, in the severe noise scenario (DA$_3$), semantic precision and recall rebound from 0.67 to 0.86, and filter recall recovers from 0.80 to 0.95. Similarly, in the fixed scaling scenario (DA$_1$), filter recall is restored from 0.56 to 0.93. Crucially, this robustness on the target domain is achieved with small or no changes in source-domain performance. The DA model maintains precision and recall scores on the source test set that are highly competitive with the baseline, confirming that the adversarial branch successfully encourages a domain-invariant latent space without discarding the critical physics information necessary for accurate hit filtering and semantic classification.

\begin{figure}[htbp!]
    \centering
    \includegraphics[width=0.7\linewidth,trim=50 0 10 10,clip]{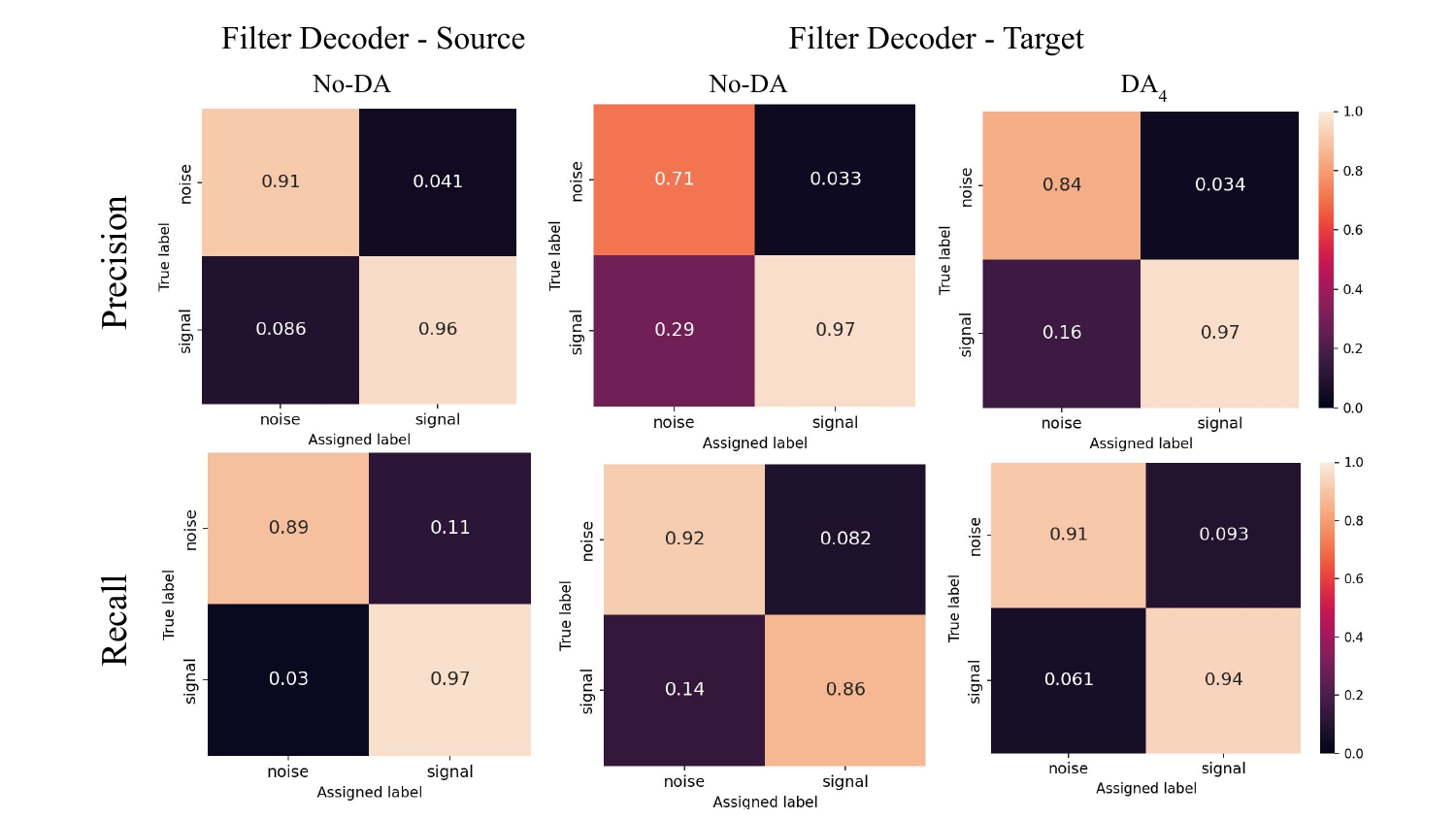}
    \caption{Same as Figure~\ref{fig:CMs_semantic}, but for filter decoder.}
    \label{fig:CMs_filter}
\end{figure}

In Figure~\ref{fig:CMs_semantic} and Figure~\ref{fig:CMs_filter} we show confusion matrices showing the performance of semantic and filter decoders, respectively, in case of drift-time-dependent RMS shift ($\mathrm{DA}_4$). On both figures, we show the performance of the model trained without DA on the source domain test set (left), on the target domain test set (middle), and performance of the model trained with DA on the same target domain test set (right). When looking at the diagonal elements for the semantic decoder, the degradation in the target domain is typically in the 1-10\% range, but DA is able to almost completely recover the performance loss. The impact is larger for the filter decoder, with a degradation of 20\% in terms of precision for the noise category, and 11\% in terms of recall for the signal category. For this variation, a larger impact on filter is expected as the network can leverage effects from diffusion to separate hits in time with the neutrino beam (signal) from hits detected at random times with respect to the beam (background, or noise). The size of this degradation is such that, while the training with DA significantly improves the filter performance of the target dataset, it is not able to fully reproduce the results from the source dataset. In Figure~\ref{fig:latents} in Appendix~\ref{sec:latents} we also show the latent space embeddings (using UMAP plot~\cite{mcinnes2018umap}) used by both decoders, colored by the two filter decoder classes (noise in red and signal in blue). We see that the confusion between the two classes is somewhat reduced in case of the model trained with DA, which leads to the improved performance in the target domain.

\vspace{-0.3cm}


\vspace{-0.5cm}
\section{Conclusion}

In this work, we present the first application of DA for GNN-based LArTPC reconstruction tasks in neutrino experiments. We mimic effects that can produce discrepancies between simulated and experimental data, such as miscalibrations, electronic
noise, and varying electron diffusion. For this we utilize the MicroBooNE open datasets as our source dataset, and introduce four different perturbations to create different target datasets. We show that our GNN model suffers from a drop in performance when these perturbations are introduced, but that training with DA, using DANN method, boosts both precision and recall in the target domain, without significant loss in performance on the source data.
Future work will include testing other DA methods, to potentially enable even stronger improvements. Additionally, we will also explore if DA can be used to bridge between different LArTPC experiments i.e., domain shift is caused by the differences in detector design and instrumentation of the two experiments, while the physics is the same.

\begin{ack}
\textit{Funding:} This work was produced by Fermi Forward Discovery Group, LLC under Contract No. 89243024CSC000002 with the U.S. Department of Energy, Office of Science, Office of High Energy Physics. Publisher acknowledges the U.S. Government license to provide public access under the DOE Public Access Plan DOE Public Access Plan
(\url{http://energy.gov/downloads/doe-public-access-plan}).

This work was supported by DOE FWP 24-09. The work of A. \'C. was also supported by the DOE Grant DE-SCL0000156.\\

We also acknowledge the MicroBooNE collaboration for providing the data used in this work.\\
\end{ack}

\medskip
\bibliographystyle{abbrv} 
\bibliography{references}

\newpage
\section*{NeurIPS Paper Checklist}

\begin{enumerate}

\item {\bf Claims}
    \item[] Question: Do the main claims made in the abstract and introduction accurately reflect the paper's contributions and scope?
    \item[] Answer: \answerYes{} 
    \item[] Justification: The abstract and introduction accurately reflect the paper’s contributions and scope, clearly motivating the problem, outlining the proposed approach, and summarizing the key findings supported by the experimental results.
    \item[] Guidelines:
    \begin{itemize}
        \item The answer \answerNA{} means that the abstract and introduction do not include the claims made in the paper.
        \item The abstract and/or introduction should clearly state the claims made, including the contributions made in the paper and important assumptions and limitations. A \answerNo{} or \answerNA{} answer to this question will not be perceived well by the reviewers. 
        \item The claims made should match theoretical and experimental results, and reflect how much the results can be expected to generalize to other settings. 
        \item It is fine to include aspirational goals as motivation as long as it is clear that these goals are not attained by the paper. 
    \end{itemize}

\item {\bf Limitations}
    \item[] Question: Does the paper discuss the limitations of the work performed by the authors?
    \item[] Answer: \answerYes{}{} 
    \item[] Justification: The paper clearly acknowledges the limitations of the proposed approach and highlights relevant directions for future investigation.
    \item[] Guidelines:
    \begin{itemize}
        \item The answer \answerNA{} means that the paper has no limitation while the answer \answerNo{} means that the paper has limitations, but those are not discussed in the paper. 
        \item The authors are encouraged to create a separate ``Limitations'' section in their paper.
        \item The paper should point out any strong assumptions and how robust the results are to violations of these assumptions (e.g., independence assumptions, noiseless settings, model well-specification, asymptotic approximations only holding locally). The authors should reflect on how these assumptions might be violated in practice and what the implications would be.
        \item The authors should reflect on the scope of the claims made, e.g., if the approach was only tested on a few datasets or with a few runs. In general, empirical results often depend on implicit assumptions, which should be articulated.
        \item The authors should reflect on the factors that influence the performance of the approach. For example, a facial recognition algorithm may perform poorly when image resolution is low or images are taken in low lighting. Or a speech-to-text system might not be used reliably to provide closed captions for online lectures because it fails to handle technical jargon.
        \item The authors should discuss the computational efficiency of the proposed algorithms and how they scale with dataset size.
        \item If applicable, the authors should discuss possible limitations of their approach to address problems of privacy and fairness.
        \item While the authors might fear that complete honesty about limitations might be used by reviewers as grounds for rejection, a worse outcome might be that reviewers discover limitations that aren't acknowledged in the paper. The authors should use their best judgment and recognize that individual actions in favor of transparency play an important role in developing norms that preserve the integrity of the community. Reviewers will be specifically instructed to not penalize honesty concerning limitations.
    \end{itemize}

\item {\bf Theory assumptions and proofs}
    \item[] Question: For each theoretical result, does the paper provide the full set of assumptions and a complete (and correct) proof?
    \item[] Answer: \answerNA{} 
    \item[] Justification: The paper does not include new theoretical results.
    \item[] Guidelines:
    \begin{itemize}
        \item The answer \answerNA{} means that the paper does not include theoretical results. 
        \item All the theorems, formulas, and proofs in the paper should be numbered and cross-referenced.
        \item All assumptions should be clearly stated or referenced in the statement of any theorems.
        \item The proofs can either appear in the main paper or the supplemental material, but if they appear in the supplemental material, the authors are encouraged to provide a short proof sketch to provide intuition. 
        \item Inversely, any informal proof provided in the core of the paper should be complemented by formal proofs provided in appendix or supplemental material.
        \item Theorems and Lemmas that the proof relies upon should be properly referenced. 
    \end{itemize}

    \item {\bf Experimental result reproducibility}
    \item[] Question: Does the paper fully disclose all the information needed to reproduce the main experimental results of the paper to the extent that it affects the main claims and/or conclusions of the paper (regardless of whether the code and data are provided or not)?
    \item[] Answer: \answerYes{} 
    \item[] Justification: The paper provides sufficient detail on the datasets, experimental setup, model configuration, training procedure, and evaluation methodology to support reproduction of the main results and conclusions.
    \item[] Guidelines:
    \begin{itemize}
        \item The answer \answerNA{} means that the paper does not include experiments.
        \item If the paper includes experiments, a \answerNo{} answer to this question will not be perceived well by the reviewers: Making the paper reproducible is important, regardless of whether the code and data are provided or not.
        \item If the contribution is a dataset and\slash or model, the authors should describe the steps taken to make their results reproducible or verifiable. 
        \item Depending on the contribution, reproducibility can be accomplished in various ways. For example, if the contribution is a novel architecture, describing the architecture fully might suffice, or if the contribution is a specific model and empirical evaluation, it may be necessary to either make it possible for others to replicate the model with the same dataset, or provide access to the model. In general. releasing code and data is often one good way to accomplish this, but reproducibility can also be provided via detailed instructions for how to replicate the results, access to a hosted model (e.g., in the case of a large language model), releasing of a model checkpoint, or other means that are appropriate to the research performed.
        \item While NeurIPS does not require releasing code, the conference does require all submissions to provide some reasonable avenue for reproducibility, which may depend on the nature of the contribution. For example
        \begin{enumerate}
            \item If the contribution is primarily a new algorithm, the paper should make it clear how to reproduce that algorithm.
            \item If the contribution is primarily a new model architecture, the paper should describe the architecture clearly and fully.
            \item If the contribution is a new model (e.g., a large language model), then there should either be a way to access this model for reproducing the results or a way to reproduce the model (e.g., with an open-source dataset or instructions for how to construct the dataset).
            \item We recognize that reproducibility may be tricky in some cases, in which case authors are welcome to describe the particular way they provide for reproducibility. In the case of closed-source models, it may be that access to the model is limited in some way (e.g., to registered users), but it should be possible for other researchers to have some path to reproducing or verifying the results.
        \end{enumerate}
    \end{itemize}

\item {\bf Open access to data and code}
    \item[] Question: Does the paper provide open access to the data and code, with sufficient instructions to faithfully reproduce the main experimental results, as described in supplemental material?
    \item[] Answer: \answerYes 
    \item[] Justification: We provide all details needed to reproduce our results. The link to both data and model code have been provided.
    \item[] Guidelines:
    \begin{itemize}
        \item The answer \answerNA{} means that paper does not include experiments requiring code.
        \item Please see the NeurIPS code and data submission guidelines (\url{https://neurips.cc/public/guides/CodeSubmissionPolicy}) for more details.
        \item While we encourage the release of code and data, we understand that this might not be possible, so \answerNo{} is an acceptable answer. Papers cannot be rejected simply for not including code, unless this is central to the contribution (e.g., for a new open-source benchmark).
        \item The instructions should contain the exact command and environment needed to run to reproduce the results. See the NeurIPS code and data submission guidelines (\url{https://neurips.cc/public/guides/CodeSubmissionPolicy}) for more details.
        \item The authors should provide instructions on data access and preparation, including how to access the raw data, preprocessed data, intermediate data, and generated data, etc.
        \item The authors should provide scripts to reproduce all experimental results for the new proposed method and baselines. If only a subset of experiments are reproducible, they should state which ones are omitted from the script and why.
        \item At submission time, to preserve anonymity, the authors should release anonymized versions (if applicable).
        \item Providing as much information as possible in supplemental material (appended to the paper) is recommended, but including URLs to data and code is permitted.
    \end{itemize}

\item {\bf Experimental setting/details}
    \item[] Question: Does the paper specify all the training and test details (e.g., data splits, hyperparameters, how they were chosen, type of optimizer) necessary to understand the results?
    \item[] Answer: \answerYes{} 
    \item[] Justification: The paper provides the relevant necessary to understand and reproduce the results.
    \item[] Guidelines:
    \begin{itemize}
        \item The answer \answerNA{} means that the paper does not include experiments.
        \item The experimental setting should be presented in the core of the paper to a level of detail that is necessary to appreciate the results and make sense of them.
        \item The full details can be provided either with the code, in appendix, or as supplemental material.
    \end{itemize}

\item {\bf Experiment statistical significance}
    \item[] Question: Does the paper report error bars suitably and correctly defined or other appropriate information about the statistical significance of the experiments?
    \item[] Answer: \answerYes{} 
    \item[] Justification: The main result is supported by an estimate of the statistical significance of the reported values, based on independent repetitions of the experiments using different random seeds.
    \item[] Guidelines:
    \begin{itemize}
        \item The answer \answerNA{} means that the paper does not include experiments.
        \item The authors should answer \answerYes{} if the results are accompanied by error bars, confidence intervals, or statistical significance tests, at least for the experiments that support the main claims of the paper.
        \item The factors of variability that the error bars are capturing should be clearly stated (for example, train/test split, initialization, random drawing of some parameter, or overall run with given experimental conditions).
        \item The method for calculating the error bars should be explained (closed form formula, call to a library function, bootstrap, etc.)
        \item The assumptions made should be given (e.g., Normally distributed errors).
        \item It should be clear whether the error bar is the standard deviation or the standard error of the mean.
        \item It is OK to report 1-sigma error bars, but one should state it. The authors should preferably report a 2-sigma error bar than state that they have a 96\% CI, if the hypothesis of Normality of errors is not verified.
        \item For asymmetric distributions, the authors should be careful not to show in tables or figures symmetric error bars that would yield results that are out of range (e.g., negative error rates).
        \item If error bars are reported in tables or plots, the authors should explain in the text how they were calculated and reference the corresponding figures or tables in the text.
    \end{itemize}

\item {\bf Experiments compute resources}
    \item[] Question: For each experiment, does the paper provide sufficient information on the computer resources (type of compute workers, memory, time of execution) needed to reproduce the experiments?
    \item[] Answer: \answerYes{} 
    \item[] Justification: We provide information about the compute we used and the time needed to run all experiments.
    \item[] Guidelines:
    \begin{itemize}
        \item The answer \answerNA{} means that the paper does not include experiments.
        \item The paper should indicate the type of compute workers CPU or GPU, internal cluster, or cloud provider, including relevant memory and storage.
        \item The paper should provide the amount of compute required for each of the individual experimental runs as well as estimate the total compute. 
        \item The paper should disclose whether the full research project required more compute than the experiments reported in the paper (e.g., preliminary or failed experiments that didn't make it into the paper). 
    \end{itemize}
    
\item {\bf Code of ethics}
    \item[] Question: Does the research conducted in the paper conform, in every respect, with the NeurIPS Code of Ethics \url{https://neurips.cc/public/EthicsGuidelines}?
    \item[] Answer: \answerYes{} 
    \item[] Justification: We have reviewed and we adhere to the NeurIPS Code of Ethics.
    \item[] Guidelines:
    \begin{itemize}
        \item The answer \answerNA{} means that the authors have not reviewed the NeurIPS Code of Ethics.
        \item If the authors answer \answerNo, they should explain the special circumstances that require a deviation from the Code of Ethics.
        \item The authors should make sure to preserve anonymity (e.g., if there is a special consideration due to laws or regulations in their jurisdiction).
    \end{itemize}

\item {\bf Broader impacts}
    \item[] Question: Does the paper discuss both potential positive societal impacts and negative societal impacts of the work performed?
    \item[] Answer: \answerNo{} 
    \item[] Justification: We do not anticipate direct negative societal impact of our work. This work will positively impact high-energy physics and neutrino research, and this topic was discussed in the manuscript.
    \item[] Guidelines:
    \begin{itemize}
        \item The answer \answerNA{} means that there is no societal impact of the work performed.
        \item If the authors answer \answerNA{} or \answerNo, they should explain why their work has no societal impact or why the paper does not address societal impact.
        \item Examples of negative societal impacts include potential malicious or unintended uses (e.g., disinformation, generating fake profiles, surveillance), fairness considerations (e.g., deployment of technologies that could make decisions that unfairly impact specific groups), privacy considerations, and security considerations.
        \item The conference expects that many papers will be foundational research and not tied to particular applications, let alone deployments. However, if there is a direct path to any negative applications, the authors should point it out. For example, it is legitimate to point out that an improvement in the quality of generative models could be used to generate Deepfakes for disinformation. On the other hand, it is not needed to point out that a generic algorithm for optimizing neural networks could enable people to train models that generate Deepfakes faster.
        \item The authors should consider possible harms that could arise when the technology is being used as intended and functioning correctly, harms that could arise when the technology is being used as intended but gives incorrect results, and harms following from (intentional or unintentional) misuse of the technology.
        \item If there are negative societal impacts, the authors could also discuss possible mitigation strategies (e.g., gated release of models, providing defenses in addition to attacks, mechanisms for monitoring misuse, mechanisms to monitor how a system learns from feedback over time, improving the efficiency and accessibility of ML).
    \end{itemize}
    
\item {\bf Safeguards}
    \item[] Question: Does the paper describe safeguards that have been put in place for responsible release of data or models that have a high risk for misuse (e.g., pre-trained language models, image generators, or scraped datasets)?
    \item[] Answer: \answerNA{} 
    \item[] Justification: This work does not use pre-trained language models, image generators, or scraped datasets, and hence poses no such risks.
    \item[] Guidelines:
    \begin{itemize}
        \item The answer \answerNA{} means that the paper poses no such risks.
        \item Released models that have a high risk for misuse or dual-use should be released with necessary safeguards to allow for controlled use of the model, for example by requiring that users adhere to usage guidelines or restrictions to access the model or implementing safety filters. 
        \item Datasets that have been scraped from the Internet could pose safety risks. The authors should describe how they avoided releasing unsafe images.
        \item We recognize that providing effective safeguards is challenging, and many papers do not require this, but we encourage authors to take this into account and make a best faith effort.
    \end{itemize}

\item {\bf Licenses for existing assets}
    \item[] Question: Are the creators or original owners of assets (e.g., code, data, models), used in the paper, properly credited and are the license and terms of use explicitly mentioned and properly respected?
    \item[] Answer: \answerYes{} 
    \item[] Justification: Our training data are derived from publicly available simulated datasets produced by MicroBooNE collaboration. Our code also builds in part on prior and ongoing work by other researchers and collaborators. All relevant sources and contributions are properly referenced and acknowledged.

    \item[] Guidelines:
    \begin{itemize}
        \item The answer \answerNA{} means that the paper does not use existing assets.
        \item The authors should cite the original paper that produced the code package or dataset.
        \item The authors should state which version of the asset is used and, if possible, include a URL.
        \item The name of the license (e.g., CC-BY 4.0) should be included for each asset.
        \item For scraped data from a particular source (e.g., website), the copyright and terms of service of that source should be provided.
        \item If assets are released, the license, copyright information, and terms of use in the package should be provided. For popular datasets, \url{paperswithcode.com/datasets} has curated licenses for some datasets. Their licensing guide can help determine the license of a dataset.
        \item For existing datasets that are re-packaged, both the original license and the license of the derived asset (if it has changed) should be provided.
        \item If this information is not available online, the authors are encouraged to reach out to the asset's creators.
    \end{itemize}

\item {\bf New assets}
    \item[] Question: Are new assets introduced in the paper well documented and is the documentation provided alongside the assets?
    \item[] Answer: \answerYes{} 
    \item[] Justification: All details about the data and the code we produced were documented and clearly presented in the manuscript.
    \item[] Guidelines:
    \begin{itemize}
        \item The answer \answerNA{} means that the paper does not release new assets.
        \item Researchers should communicate the details of the dataset\slash code\slash model as part of their submissions via structured templates. This includes details about training, license, limitations, etc. 
        \item The paper should discuss whether and how consent was obtained from people whose asset is used.
        \item At submission time, remember to anonymize your assets (if applicable). You can either create an anonymized URL or include an anonymized zip file.
    \end{itemize}

\item {\bf Crowdsourcing and research with human subjects}
    \item[] Question: For crowdsourcing experiments and research with human subjects, does the paper include the full text of instructions given to participants and screenshots, if applicable, as well as details about compensation (if any)? 
    \item[] Answer: \answerNA{}
    \item[] Justification: This work does not include crowdsourcing and research with human subjects.
    \item[] Guidelines:
    \begin{itemize}
        \item The answer \answerNA{} means that the paper does not involve crowdsourcing nor research with human subjects.
        \item Including this information in the supplemental material is fine, but if the main contribution of the paper involves human subjects, then as much detail as possible should be included in the main paper. 
        \item According to the NeurIPS Code of Ethics, workers involved in data collection, curation, or other labor should be paid at least the minimum wage in the country of the data collector. 
    \end{itemize}

\item {\bf Institutional review board (IRB) approvals or equivalent for research with human subjects}
    \item[] Question: Does the paper describe potential risks incurred by study participants, whether such risks were disclosed to the subjects, and whether Institutional Review Board (IRB) approvals (or an equivalent approval/review based on the requirements of your country or institution) were obtained?
    \item[] Answer: \answerNA{} 
    \item[] Justification: This work does not involve human subjects or study participants, and therefore IRB approval or equivalent ethical review is not applicable.
    \item[] Guidelines:
    \begin{itemize}
        \item The answer \answerNA{} means that the paper does not involve crowdsourcing nor research with human subjects.
        \item Depending on the country in which research is conducted, IRB approval (or equivalent) may be required for any human subjects research. If you obtained IRB approval, you should clearly state this in the paper. 
        \item We recognize that the procedures for this may vary significantly between institutions and locations, and we expect authors to adhere to the NeurIPS Code of Ethics and the guidelines for their institution. 
        \item For initial submissions, do not include any information that would break anonymity (if applicable), such as the institution conducting the review.
    \end{itemize}

\item {\bf Declaration of LLM usage}
    \item[] Question: Does the paper describe the usage of LLMs if it is an important, original, or non-standard component of the core methods in this research? Note that if the LLM is used only for writing, editing, or formatting purposes and does \emph{not} impact the core methodology, scientific rigor, or originality of the research, declaration is not required.
    \item[] Answer: \answerNA{} 
    \item[] Justification: LLMs were used only for editing and formatting assistance and were not used as part of the core methodology, experimental design, analysis, or scientific conclusions.
    \item[] Guidelines:
    \begin{itemize}
        \item The answer \answerNA{} means that the core method development in this research does not involve LLMs as any important, original, or non-standard components.
        \item Please refer to our LLM policy in the NeurIPS handbook for what should or should not be described.
    \end{itemize}

\end{enumerate}

\newpage
\appendix
\section{Latent Space Visualization}\label{sec:latents}
\begin{figure}[htbp!]
    \centering
    \includegraphics[width=0.99\linewidth,trim=80 20 35 30,clip]{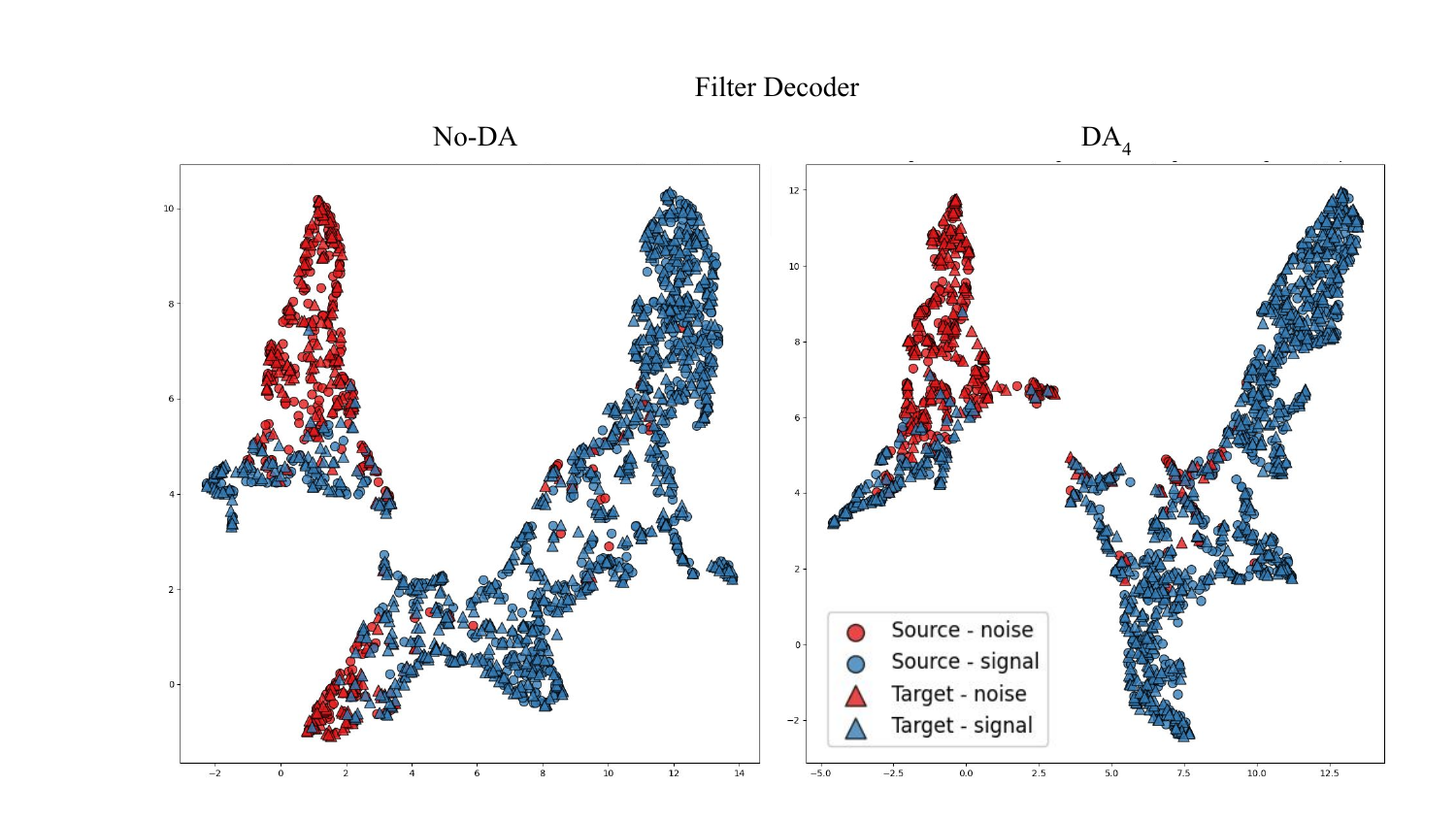}
    \caption{Latent space embeddings (plotted using UMAP~\cite{mcinnes2018umap} plot with 10 neighbors for each point) for models trained without (left) and with DA (right), in $\mathrm{DA}_4$ experiments. Source domain (circles), and target domain (triangles) data with two filter decoder classes (noise in red, and signal in blue) are shown. Inclusion of DA improves alignment and helps slightly improve separation between the two classes.}
    \label{fig:latents}
\end{figure}

\end{document}